# CyberMoraba: A game-based approach enhancing cybersecurity awareness

Mike Nkongolo

Mike.wankongolo@up.ac.za

University of Pretoria, Faculty of Informatics, South Africa

## Abstract

Numerous studies have confirmed the effectiveness of Cybersecurity Awareness Games (CAGs) in enhancing the security posture of diverse organisations. As these organisations increasingly face the formidable challenge of cyberattacks, implementing serious CAGs to solve this issue has become a paramount concern. This article introduces an innovative approach to cybersecurity education by presenting a serious CAG. The game aims to effectively educate students about critical aspects of cybersecurity awareness engagingly and interactively. The study aimed to redefine cybersecurity awareness training by introducing an indigenous game design that intricately incorporates the traditional South African *Morabaraba* board game. While the effectiveness of non-indigenous games like "Capture The Flag (CTF)" in cybersecurity training is acknowledged, indigenous designs have been overlooked. This research creatively integrates Morabaraba's gameplay into cybersecurity training, adapting it into a competitive game where players adopt the roles of either defenders or attackers, with corresponding tokens/images symbolising various cyber defense and attack strategies. Both the defenders and attackers in the game can elevate their awareness scores by strategically positioning defensive or attacking images on the game board. Subsequently, a judging entity assesses the players' moves and assigns scores based on the accuracy of the images placed. The game mirrors real-world scenarios, promoting strategic thinking and leveraging interactive gameplay for practical insights into cybersecurity awareness. Players demonstrate their cybersecurity knowledge through offensive and defensive strategies. A group of 40 students evaluated the game's effectiveness, highlighting its potential to create an engaging and competitive learning environment that imparts cybersecurity principles and practical application. The evaluation of the game mechanics demonstrated a remarkably positive outcome, with participants expressing both enjoyment and an enhanced understanding of cybersecurity awareness.

## 1. Introduction

Cybersecurity awareness refers to the knowledge, understanding, and vigilance that individuals, organisations, and communities possess regarding potential cybersecurity threats, risks, and best practices (Abd et al, 2015). It involves being informed about various cyber threats, understanding how to protect sensitive information, and adopting proactive measures to prevent cyberattacks. Cybersecurity awareness is vital for safeguarding individuals and organisations from a wide range of cyber threats, including viruses, malware, phishing, and ransomware (Chowdhury & Gkioulos, 2021; Abd et al, 2015). It aids in preventing data breaches, financial losses, and reputational damage, while also educating about personal privacy protection and secure online practices. Moreover, it ensures compliance with industry regulations, defends against social engineering tactics, and safeguards critical infrastructure from potential disruptions (Naidoo & Jacobs, 2023; Nkongolo & Tokmak, 2023). Embracing cybersecurity awareness reflects a global responsibility to maintain a secure digital landscape, fostering responsible online behavior and contributing to a safer internet for everyone. The impact of comprehensive cybersecurity awareness training and education is profound. When individuals are well-trained and educated about cybersecurity, their ability to recognise, prevent, and respond to cyber threats significantly improves. This leads to a strengthened overall security posture for both individuals and organisations (Abd et al, 2015). Informed individuals are better equipped to identify phishing attempts, avoid malicious links, protect sensitive information, and practice secure online behaviors (Chowdhury & Gkioulos, 2021). Moreover, they contribute to creating a culture of cybersecurity, where knowledge is shared and best practices are followed, enhancing the collective resilience against cyberattacks. Ultimately, investing in cybersecurity awareness training results in a more vigilant, proactive, and security-conscious digital community (Chowdhury & Gkioulos, 2021; Naidoo & Jacobs, 2023). In this paper, we present "*CyberMoraba*," a game meticulously crafted to educate students about the nuances of cybersecurity awareness. Numerous games have been suggested to aid in combating cybercrime. Prominent games encompass *CyberCIEGE*, which adeptly simulates real-world cyber incidents and furnishes players with a platform to hone their defense against cyber threats (Irvine et al, 2005). Similarly, *CTRL-ALT-Hack* immerses players in simulated hacking scenarios, fostering a hacker's perspective to bolster defenses against cyber threats (Kipker et al, 2021). *CyberStart* ingeniously imparts cybersecurity skills to students, propelling them to conquer challenges and puzzles (Stoker et al, 2021), while *R00tz Asylum* intricately weaves interactive games into the domain of young individuals, imparting knowledge about hacking, coding, and cybersecurity (Haroldson & Ballard, 2021). Further, Cybersecurity *Capture The Flag (CTF)* stands as an engaging challenge where participants decode cybersecurity puzzles and execute tasks to amass points (Ortiz-Garces et al, 2023). Nonetheless, the distinguishing aspect among these games lies in their omission of indigenous game design. This observation underscores the untapped potential within the realm of cybersecurity awareness training. The incorporation of indigenous game design offers a distinct avenue for imparting knowledge and understanding, aligning with cultural sensibilities and traditional narratives. By embracing indigenous game elements, a unique bridge is formed between cybersecurity education and heritage, fostering a more resonant and relatable approach to raising awareness. This symbiosis not only introduces fresh perspectives to cybersecurity but also enriches the fabric of educational experiences by harnessing the power of cultural familiarity. To the best of our knowledge, there has been no instance of a serious CAG being developed with an indigenous gameplay such as *Morabaraba*. Morabaraba, also known as "*Umlabalaba*," is a traditional two-player strategy board game with African origins (Nkopodi & Mosimege, 2009; Nkongolo, 2023), particularly prevalent among the Bantu-speaking peoples of South Africa. The game is played on a distinctive board with three concentric squares, each comprised of eight intersection points (Nkopodi & Mosimege, 2009; Nkongolo, 2023). Players have twelve pieces, typically represented by cowrie shells or stones, and the objective is to strategically move these pieces to form specific patterns and capture the opponent's pieces (Nkopodi & Mosimege, 2009; Nkongolo, 2023). Morabaraba is known for its intricate gameplay, requiring careful planning, tactical thinking, and adaptability (Nkopodi & Mosimege, 2009; Nkongolo, 2023). It has cultural and social significance within various African communities, serving as a means of entertainment and often symbolising intelligence and strategic prowess (Nkongolo, 2023). It is a traditional two-player strategy board game that holds cultural significance in the region (Nkopodi & Mosimege, 2009; Nkongolo, 2023). The game is characterised by a distinctive game board layout and involves players attempting to align their game pieces to form rows, thereby gaining an advantage over their opponent (Nkongolo, 2023). It has historical roots within indigenous communities and continues to be played in both traditional and modern settings (Nkopodi & Mosimege, 2009). The Morabaraba game's strategic elements offer innovative avenues to elevate cybersecurity awareness training. The gameplay of Morabaraba is intricate and strategic. Two players, each with twelve pieces, take turns to make their moves on the board, aiming to create specific patterns and capture their opponent's pieces (Nkopodi & Mosimege, 2009). The game progresses in two phases:

**Placement phase**: In this initial phase, players take turns placing their twelve pieces on the board's intersection points, one piece at a time (Nkopodi & Mosimege, 2009). The goal is to strategically position the pieces for future movement and pattern creation.

**Movement phase:** Once all pieces are placed, the game transitions into the movement phase. During this phase, players take turns moving one of their pieces along the lines on the board, attempting to form specific patterns (Nkopodi & Mosimege, 2009). The patterns include creating a "Mill," which consists of three of their pieces in a straight line, horizontally or vertically (Nkopodi & Mosimege, 2009). When a player forms a Mill, they can capture one of their opponent's pieces. The game continues until one player can no longer make a legal move or is reduced to only three pieces (Nkopodi & Mosimege, 2009), which allows the opponent to capture their pieces freely. The player who captures all their opponent's pieces or immobilises them wins the game. Morabaraba's gameplay requires careful planning, strategic thinking, and adaptability, making it a challenging and engaging traditional board game (Nkopodi & Mosimege, 2009; Nkongolo, 2023). Integrating Morabaraba's strategic dynamics into cybersecurity education not only enhances participants' understanding but also makes training engaging and relatable. This approach effectively combines traditional game principles with modern security challenges, fostering practical skills acquisition and bridging the gap between entertainment and education in the realm of cybersecurity. The research question is formulated as follows:

*How does the integration of traditional South African Morabaraba gameplay into cybersecurity training, as exemplified by CyberMoraba, impact the effectiveness of cybersecurity awareness and practical application among students?*

This research question explores the influence of introducing Morabaraba-inspired elements into cybersecurity training, using CyberMoraba as a model. It focuses on assessing how this unique approach affects students' cybersecurity awareness and their ability to practically apply cybersecurity principles. The question aims to measure the educational value and effectiveness of such an indigenous game design in enhancing cybersecurity knowledge and skills. The target group for the CyberMoraba game is a critical aspect of its design and implementation. It is not a one-size-fits-all approach, as different groups within an organisation may have varying levels of cybersecurity awareness and different roles and responsibilities in ensuring cybersecurity (Abd et al, 2015; Ortiz-Garces et al, 2023; Naidoo & Jacobs, 2023). For this study, the primary target audience is the students, and it is essential to customize the game's usage based on their specific needs. However, within an organisational context, various personnel groups, such as HR (Human Resources), finance, and technical personnel, may have unique cybersecurity challenges and requirements, and the game can be adapted to address these distinct aspects (Ortiz-Garces et al, 2023). In essence, the game's flexibility and versatility are key strengths, allowing organisations to tailor its use to specific target groups, and ensuring that it remains relevant and impactful across various departments and individuals within the organisation.

## 2. Methodology

In this section, we illuminate various facets of the proposed game's design. We commence by outlining the formulation of the game's rules, subsequently delving into an elucidation of the constituents that constitute the game environment, serving as the backdrop for the gameplay itself. The discourse then concludes by elaborating upon the game's scoring mechanics and the technological architecture underpinning the *Graphical User Interface (GUI)*. The components that comprise the game GUI are visually depicted in Figure 1.

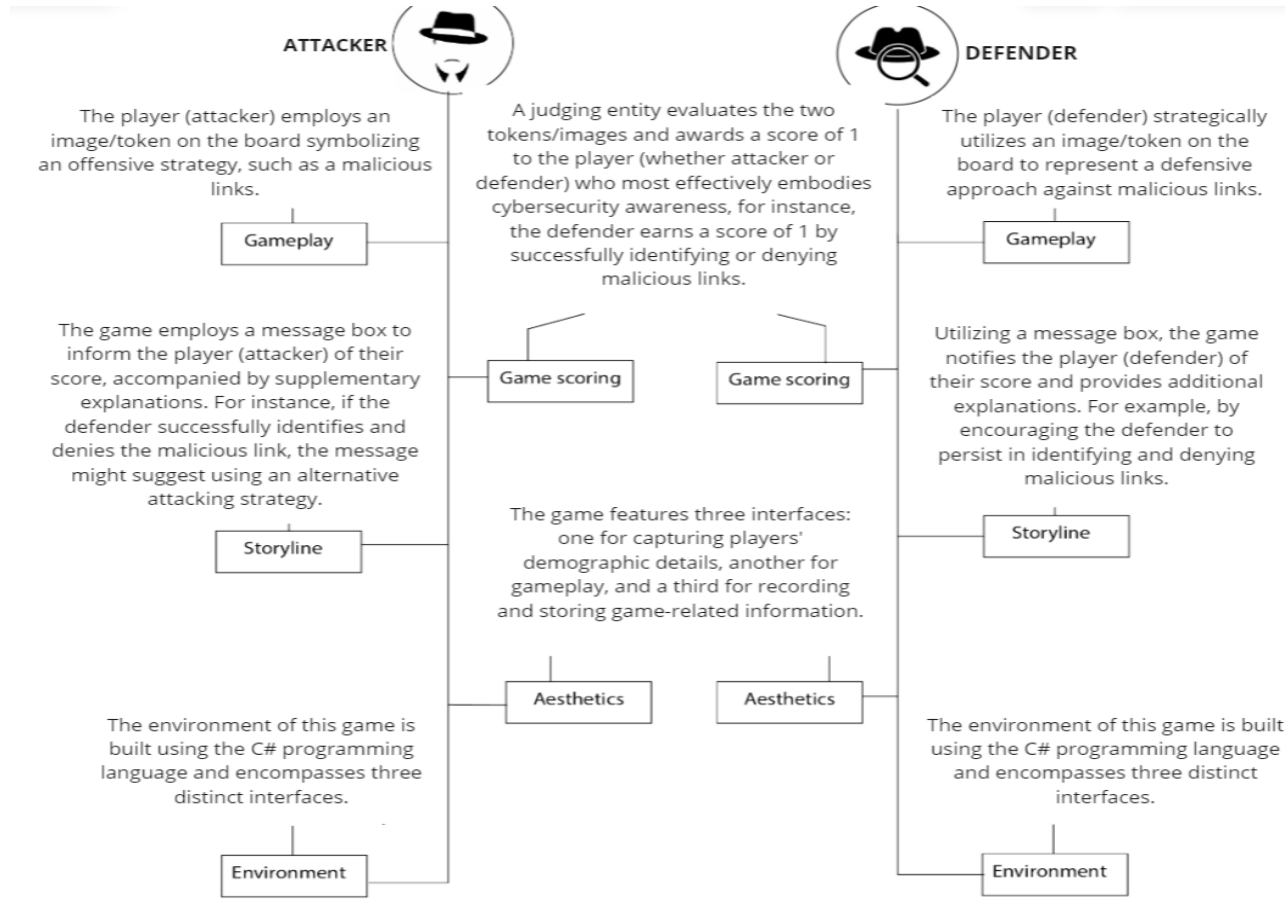

**Figure 1.** The game components

### 2.1 Rules of the game

Each player is given a set of 13 tokens or images that represent attacking or defending strategies (Table 1). The attacker's tokens are labelled with A1, A2, A3, …, A13, and the defender's tokens are labelled with D1, D2, D3, …, D13 (Table 1). During the game, each player takes turns placing their tokens on the game board, with each token placement representing a move that determines whether the attacker or defender used an optimal strategy (Nkongolo, 2023). Commencing with an unoccupied board, the game's initiation rests with the attacker. Subsequently, the judging entity bestows a positive reward of 1 upon the player executing the most adept move, while assigning a zero to the player opting for an incorrect move (Traulsen & Glynatsi, 2023). Progress within the game materialises through the placement of the appropriate tokens onto the board. The game format is designed to simulate real-world cybersecurity scenarios and help a person gain a deeper understanding of cybersecurity best practices (Abd et al, 2015). The game's clear rules and sequential nature make it easy for players to understand and engage with, while the judging entity motivates players to perform their best. The strategies of the game require players to think strategically about their moves and anticipate their opponent's next move (Nkongolo, 2023).

Successful players need to balance their offensive and defensive moves to gain an advantage on the board. The objective of the game is to simulate a scenario in which the defender and attacker compete to accomplish their respective objectives (Traulsen & Glynatsi, 2023). The defender aims to obtain the highest cybersecurity awareness score by either correctly placing the appropriate image on the board or using the most efficient cybersecurity strategy to defend against the attacker's actions. Conversely, the attacker strives to achieve the highest intrusion score by implementing the most effective attacking strategy to penetrate the defender's cybersecurity defense.

**Table 1.** The attacker and defender tokens

| Token | Image | Definition | Token | Image | Definition |
|---|---|---|---|---|---|
| A1 | Email | Malicious e-mail | D1 | Denying | Blocking/denying actions |
| A2 | Phone | Malicious phone call | D2 | Network monitoring | Network traffic analysis |
| A3 | Chat | Malicious chat | D3 | Avoid clicking | Refuse to click |
| A4 | Attachment | Malicious attachment | D4 | Identification | Verification process |
| A5 | Donate | Malicious directory | D5 | No trust | Zero trust policy |
| A6 | Password | Malicious directory | D6 | Upload | Uploading process |
| A7 | Connection | Malicious network connection | D7 | Trust | The defender trusts |
| A8 | Access | Malicious intrusion | D8 | Provide | Providing information |
| A9 | Data | Malicious data | D9 | Confidential | Confidentiality of data |
| A10 | Data loss | Data loss process | D10 | Report | Reporting cyber incidents |
| A11 | Click | Malicious link | D11 | Social media | Sharing data on social media |
| A12 | Sensitive data | Theft of data | D12 | Connection | Secured network connection |
| A13 | Message | Malicious communication | D13 | Backup | Data recovery |

Winning the game necessitates that the defender understands the various cybernetic strategies that the attacker may utilise and can select the optimal defense approach to combat each one. The attacker must leverage their knowledge of data protection vulnerabilities to use the correct image to score points. In the same vein, the defender must be watchful and put the correct image on the board to thwart the attacker's successful attacks. To win the game, it is essential to have a comprehensive data protection strategy that includes both technical and non-technical measures. To compute the final score of each player when the game is over, the judging entity evaluates the tokens as illustrated in Table 2 where A% and D% represent the attacker and defender score (Traulsen & Glynatsi, 2023; Nkongolo, 2023). It is important to note that the types and number of images can be expanded. Figure 1 and Figure 2 present the attacker and defender tokens/images (Figure 3). The game ends when all 13 images are placed on the board. If both players have the same total score, the game is drawn.

**2.2 Architecture of the game**

The game's software was developed using Visual Studio Integrated Development Environment (IDE) and necessitates a desktop or personal computer equipped with a connected mouse and keyboard to facilitate gameplay (Traulsen & Glynatsi, 2023). Elevating the game's challenge can be achieved by incorporating a timer mechanism. In the game's sequence of play, the player initiates by submitting demographic details and choosing their preferred role, either attacker or defender (Figure 4). Subsequently, they access the game board through a menu, where they select images to signify their strategic moves (Figure 5 and Figure 6).

**Table 2.** An instance of the judging scoring scheme

|   | A | D | A% | D% | Judge | Game Feedback/Message |
|---|---|---|---|---|---|---|
| 1 | Email | Zero trust | 0 | 1 | Defender best move | Never trust malicious emails |
| 2 | Click | Denying | 0 | 1 | Defender best move | Keep denying malicious links |
| 3 | Chat | Identification | 0 | 1 | Defender best move | Identification of malicious chats |
| 4 | Phone call | Trust | 1 | 0 | Attacker best move | The defender trusted a malicious call |
| 5 | Connection | Connection | 0 | 1 | Defender best move | Secured connection suggested |
| 6 | Access | Identification | 0 | 1 | Defender best move | Malicious access identified |
| 7 | Data loss | Upload | 1 | 0 | Attacker best move | Data loss occurred |
| 8 | Click | Provide | 1 | 0 | Attacker best move | Malicious link used |
|   | **Total** |  | 3% | 5% | Defender wins with 5% | **Defender best moves**: *zero trust, denying, identification, secured connection* |

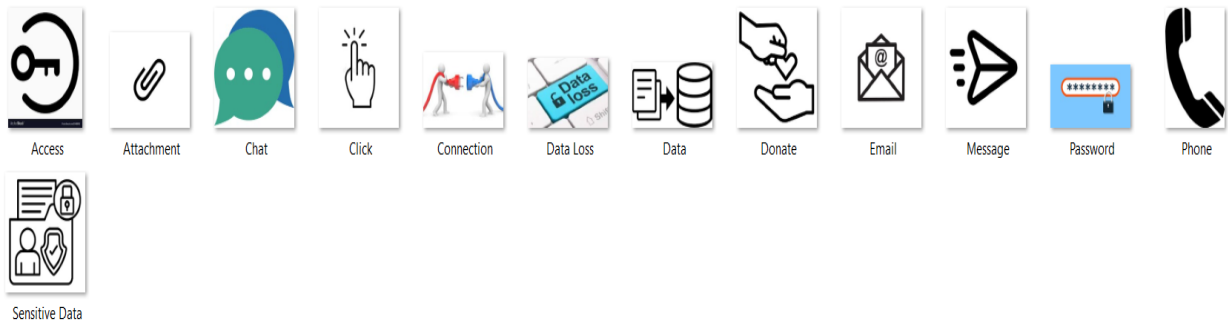

**Figure 2.** The attacker tokens (author's images)

In the left depiction of Figure 6, the attacker successfully convinces the defender to access a malicious website, leading to the compromise of the defender's credentials. As a result, the attacker achieves a score of 1. In the right depiction of Figure 6, the attacker proposes a deceitful connection, met with the defender's skepticism. The defender recommends a secure alternative, showcasing resilience, and earns a score of 1. Once all images have been utilised by both players, the game concludes, and the record menu will save players' information (Figure 7). The scoring mechanism employed by the judging entity is implemented with a conditional "if-else" structure based on Table 2.

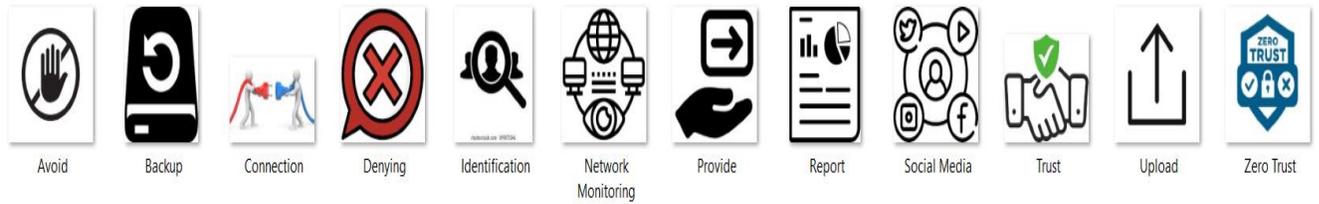

**Figure 3.** The defender tokens (author's images)

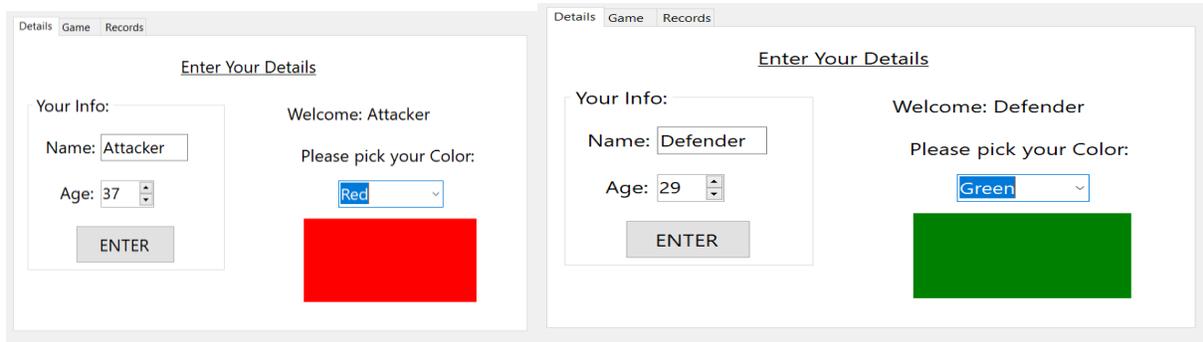

**Figure 4.** The player role selection

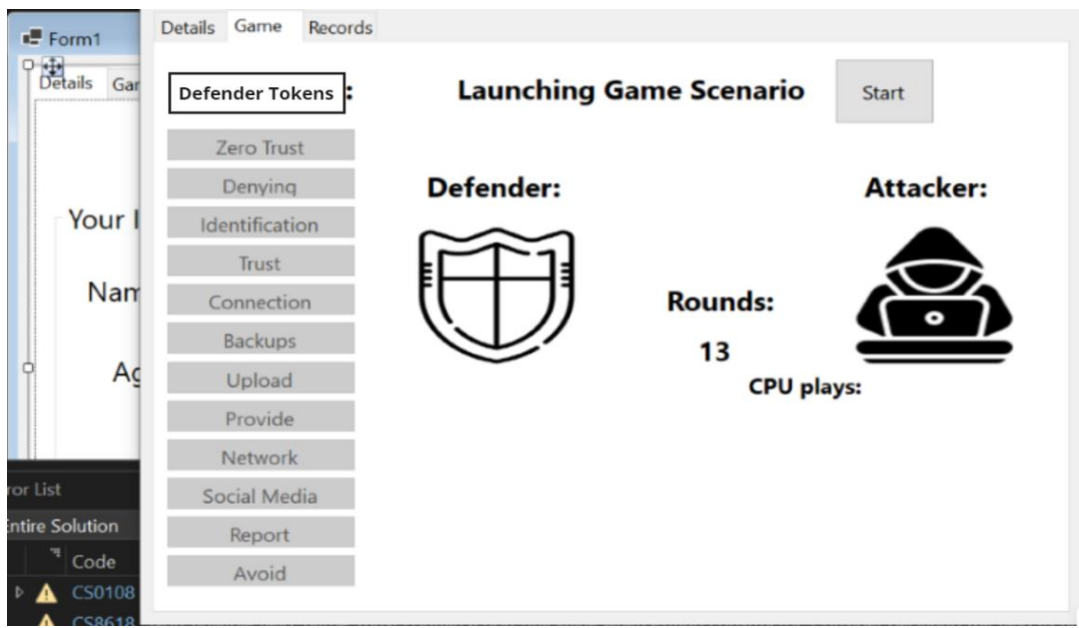

**Figure 5.** Launching the gameplay scenario

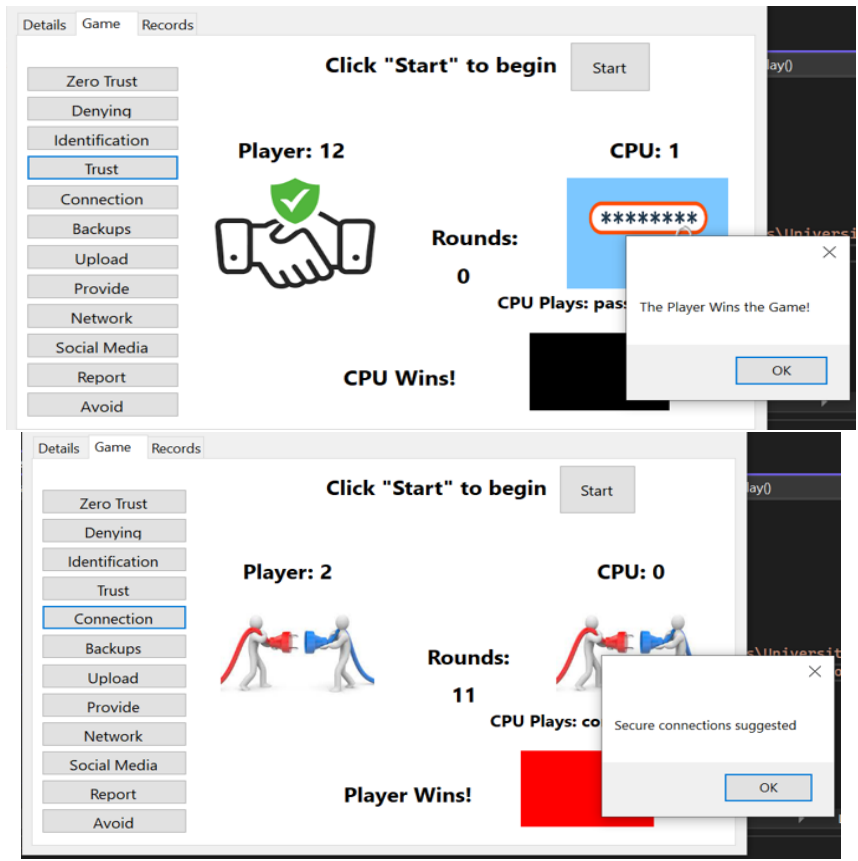

**Figure 6.** Illustration of a scenario involving game mechanics

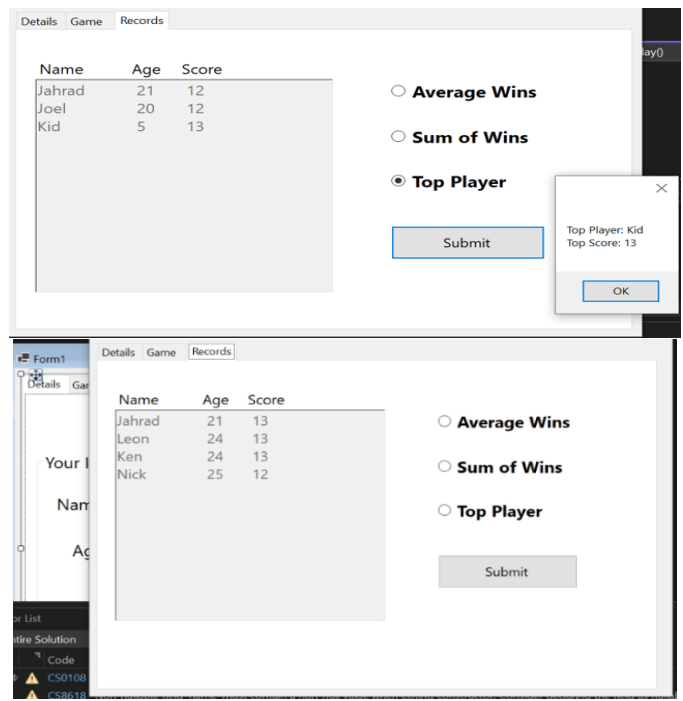

**Figure 7.** Illustration of recorded players' information

## 3. Results of players' feedback assessment

For this research, a cohort of 40 student participants was meticulously chosen from the university community, encompassing diverse backgrounds and possessing a commendable familiarity with gaming dynamics as well as a solid understanding of the realm of cybersecurity awareness. Participants were not allowed to specify improvement areas in the questions provided. Additionally, questions related to the indigenous game design element were not included in the questionnaire. The survey administered to the participants encompassed the following inquiries:

**On a scale of 0 to 10, with 0 indicating unlikeliness and 10 signifying utmost likelihood, how probable would you be to engage in this gameplay?**

This question seeks to understand students' level of interest and willingness to engage in gameplay. The findings from this question can provide insights into the overall appeal of the game. If most students rate their likelihood to engage at higher values (7-10), it suggests the gameplay has garnered significant interest (Figure 8 and Figure 9).

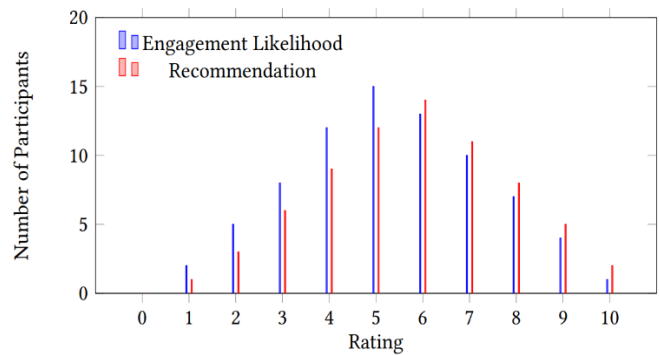

**Figure 8.** Participants gameplay interest

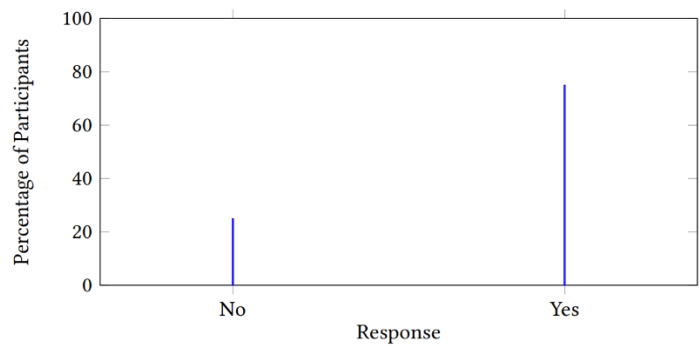

**Figure 9.** Participants gameplay responses

Conversely, lower ratings (0-6) might indicate areas where the game could be improved to increase engagement (Figure 8).

**Would you consider recommending this game to others for their participation?**

This question delves into students' willingness to recommend the game to others. Positive responses (Yes) indicate that students found the game enjoyable and valuable enough to recommend to their peers (Figure 10).

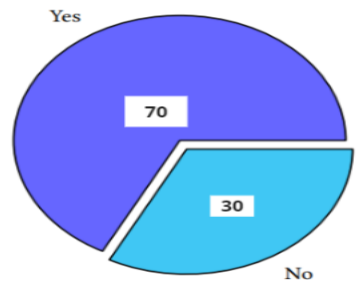

**Figure 10.** Perceived potential of the game to educate

Conversely, negative responses (No) might imply that students identified areas where the game could be enhanced before they would recommend it.

**What aspects of this game would you find enjoyable and engaging?**

Figure 11 indicates that students found various aspects of the game enjoyable and engaging, with the highest percentage leaning toward gameplay, followed by challenges and interactive elements. This information can help guide future game design decisions and improvements to enhance the overall engagement and effectiveness of the game in promoting cybersecurity awareness.

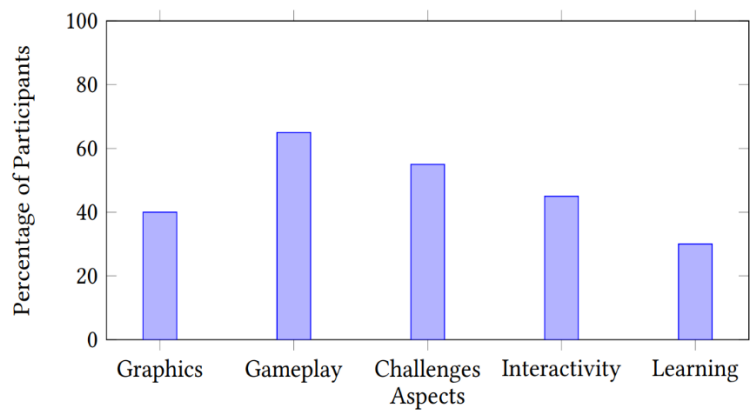

**Figure 11.** Participants perceptions of various aspects of the game

**In your perspective, what enhancements or modifications could be made to elevate the game's overall experience?**

Figure 12 demonstrates that students had varied opinions on how to enhance the game's overall experience. The top three categories of suggestions—More Challenges, Improved Interactivity, and Better Graphics—reflect students' desire for engaging gameplay with higher levels of complexity, interactive elements, and visual appeal. This information is valuable for informing game development decisions, aligning with participants' preferences, and crafting a more enjoyable and impactful game for promoting cybersecurity awareness.

## 4. Discussion and principal findings

The emergence of serious games within cybersecurity is rapidly expanding (Abd et al, 2015). This research delves into the intricate landscape of constructing a potent serious game designed in the context of cybersecurity awareness (Nkongolo, 2023). Through our exploration, we aim to provide readers with a comprehensive insight into harnessing indigenous game designs to fashion efficacious serious games for countering cyber threats (Chowdhury & Gkioulos, 2021). Notably, the players/students' responses indicated a strong engagement with both the gameplay and storyline—a pivotal facet. Achieving equilibrium between entertainment and education is a paramount objective for serious games, ensuring the creation of an environment conducive to effective learning (Calvano et al, 2023).

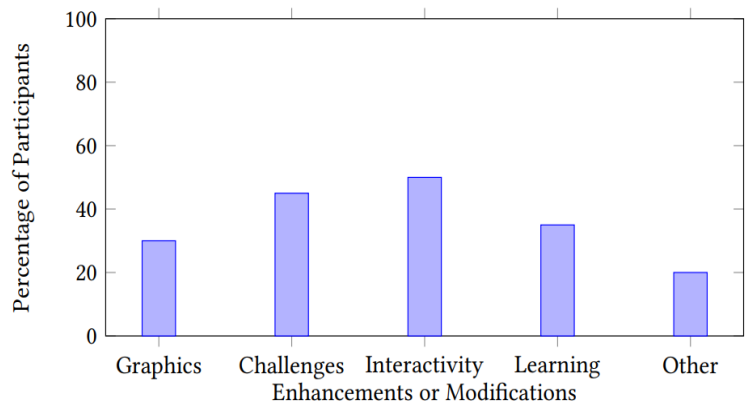

**Figure 12.** Participants' perceptions of gameplay modification

We elucidated the methodology for crafting a game that seamlessly blends enjoyment and educational value by showcasing functional mechanics. This elucidation carries significance, as a game that fails to resonate positively with players runs the risk of being ineffective as a learning tool (Moumouh et al, 2023). Consequently, the vital synergy between enjoyment and learning stands as a fundamental principle for the efficacy of serious CAGs (Traulsen & Glynatsi, 2023). The research introduces several noteworthy contributions, including innovative game mechanics, a multi-interface gameplay approach, and enhanced gameplay elements. The unique mechanics involve players/students assuming roles as defenders or attackers to address cyber threats, fostering a better understanding of cybersecurity. The game employs three C# interfaces to create a diverse and comprehensive learning environment. Additional features like timers and tokens/images enhance the gameplay, providing more challenges and learning opportunities. Named "*CyberMoraba*," the game creatively blends traditional Morabaraba gameplay with cybersecurity awareness education. *Morabaraba* and *CyberMoraba* share a fundamental similarity in their competitive and strategic gameplay. Both games are designed to engage players in strategic thinking and decision-making, making them ideal platforms for learning and skill development:

**Competitive nature:** Morabaraba, is known for its competitive gameplay (Nkopodi & Mosimege, 2009). Similarly, CyberMoraba introduces a competitive element to cybersecurity awareness training. In CyberMoraba, players take on the roles of defenders and attackers, competing to outsmart each other in a simulated cyber environment. This competitive aspect adds an element of excitement and engagement to the learning process in both games.

**Strategy and decision-making:** Morabaraba requires players to think strategically and plan their moves carefully (Nkopodi & Mosimege, 2009). The placement of tokens on the board can significantly impact the outcome of the game. CyberMoraba extends this strategic thinking to the realm of cybersecurity. Players must make informed decisions on where to position defensive and attacking images on the game board to maximize their awareness scores. This strategic element encourages critical thinking and problem-solving, a common trait in both games.

**Real-world simulation:** Both Morabaraba and CyberMoraba offer a degree of real-world simulation. In Morabaraba, players replicate the strategic thinking used in traditional cattle herding (Nkopodi & Mosimege, 2009). In CyberMoraba, the game mirrors real-world cybersecurity scenarios, allowing players to apply their knowledge practically. This real-world context is a shared feature that enhances the educational value of both games.

**Interactive learning:** Morabaraba has long been used as a means of interactive learning, teaching players valuable skills and cultural knowledge (Nkopodi & Mosimege, 2009). CyberMoraba extends this tradition by incorporating elements of cybersecurity awareness and education into its gameplay. Both games provide an interactive and engaging learning experience, making them effective tools for imparting knowledge.

**Skill development:** Morabaraba hones players' skills in strategy, planning, and decision-making (Nkopodi & Mosimege, 2009). Similarly, CyberMoraba enhances players' skills related to cybersecurity awareness, defense, and attack strategies. Both games focus on skill development within their respective domains. The relationship between Morabaraba and CyberMoraba lies in their shared emphasis on competitive gameplay, strategic thinking, real-world simulation, interactive learning, and skill development (Nkopodi & Mosimege, 2009; Nkongolo, 2023). CyberMoraba builds upon the traditional game of Morabaraba, adapting its gameplay to educate and engage students in the field of cybersecurity.

### 4.1. Further research development

Presently, our efforts are focused on the ongoing improvement of CyberMoraba by introducing multiple 3D levels to enhance its appeal among students. The logo of the CyberMoraba 3D game is shown in Figure 13.

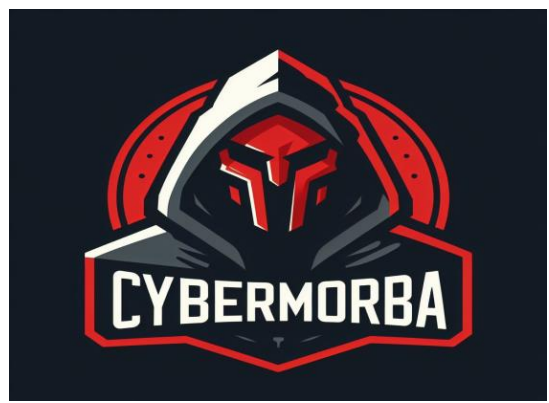

**Figure 13.** The CyberMoraba logo

This initiative aligns with our commitment to advancing Educational Serious Game Design (ESGD), underscoring our anticipation of continued endeavors in this domain. Furthermore, we envisage extending the evaluation of CyberMoraba to encompass more and various participants with different backgrounds. We also intend to create a function to capture the game outcome and playtime (Figure 14). This data will be recorded as players interact with the GUI of the game to form a dataset that includes various states of the game (Figure 14). The dataset will be used to predict players' actions using machine learning.

| Nickname | Defender Score | Attacker Score | Time (sec) | Winner |
|---|---|---|---|---|
| John | 4 | 8 | 152 | Attacker |
| Arthur | 6 | 6 | 127 | Draw |
| Tristin | 5 | 7 | 145 | Attacker |
| Jess | 6 | 6 | 82 | Draw |
| Steve | 7 | 5 | 110 | Defender |
| JP | 7 | 5 | 118 | Defender |
| Kenny | 5 | 7 | 144 | Attacker |
| TIm | 7 | 5 | 89 | Defender |
| Melissa | 2 | 10 | 102 | Attacker |

**Figure 14.** The CyberMoraba Scoreboard

### 4.2. Challenges and applicability

While the initial CyberMoraba evaluation with students yielded positive results, it is important to acknowledge the limitations of the study. One limitation is that the study primarily focused on students, and the effectiveness of CyberMoraba with other target groups, such as professionals in various organisations, remains unexplored. Generalising the game's impact beyond the student demographic should be approached with caution. Additionally, the study did not include specific questions related to the indigenous game design element, which could have provided valuable insights into its effectiveness and cultural relevance. This can be explored in future research with a distinct focus group separate from the student population. Future work directions should aim to address these limitations and broaden the game's applicability. A concrete direction for future research involves conducting similar evaluations with diverse target groups, including professionals and individuals from different cultural backgrounds. This will help determine the game's generalisability and effectiveness in various educational contexts. Moreover, further studies should incorporate questions that specifically assess the impact of the indigenous game design on the learning experience.

## 5. Conclusion

Within the scope of this research paper, we introduced a novel ESGD aimed at imparting cybersecurity awareness to students by employing the innovative Morabaraba game design. The game's architecture enabled players to adopt the roles of either defenders or attackers, with corresponding tokens/images symbolising various cyber defense and attack strategies. The initial prototype of the game was presented, followed by a comprehensive evaluation process. Student responses underscored the game's potential as a robust learning tool for cybersecurity awareness, as well as their inclination to recommend it to others. The implemented cybersecurity game effectively addresses a significant void within the African academic landscape, as it constitutes the first of its kind not only in Africa but also worldwide. Capitalising on the growing allure of video games among the digital-native generation, the integration of *CyberMoraba* as a vehicle for promoting cybersecurity awareness is poised to invigorate student motivation and enthusiasm toward acquiring essential cybersecurity skills. In the future, the discoveries made in this paper will be subject to comparison with the existing body of literature about ESGD.

## 6. Acknowledgement

The author expresses gratitude to the University of Pretoria's Faculty of Engineering, Built Environment, and Information Technology for their support in funding this research project through the Doctorate University Capacity Development Program (UCDP) Grant A1F637.

## References

Abd Rahim, N.H., Hamid, S., Kiah, M.L.M., Shamshirband, S. and Furnell, S., 2015. A systematic review of approaches to assessing cybersecurity awareness. Kybernetes, 44(4), pp.606-622.

Calvano, M., Caruso, F., Curci, A., Piccinno, A. and Rossano, V., 2023. A Rapid Review on Serious Games for Cybersecurity Education: Are "Serious" and Gaming Aspects Well Balanced? 9th International Symposium on End-User Development, 6-8 June 2023, Cagliari, Italy. CEUR Workshop Proceedings.

Chowdhury, N. and Gkioulos, V., 2021. Cyber security training for critical infrastructure protection: A literature review. Computer Science Review, 40, p.100361.

Haroldson, R. and Ballard, D., 2021. Alignment and representation in computer science: an analysis of picture books and graphic novels for K-8 students. Computer Science Education, 31(1), pp.4-29.

Irvine, C.E., Thompson, M.F. and Allen, K., 2005. CyberCIEGE: gaming for information assurance. IEEE Security & Privacy, 3(3), pp.61-64.

Kipker, D.K. and Pape, S., 2021. Case Study: Checking a Serious Security-Awareness Game for its Legal Adequacy. Datenschutz und Datensicherheit-DuD, 45, pp.310-314.

Moumouh, C., Chkouri, M.Y. and Fernández-Alemán, J.L., 2023. Cybersecurity Awareness Through Serious Games: A Systematic Literature Review. In International Conference on Networking, Intelligent Systems and Security (pp. 190-199). Springer, Cham.

Naidoo, R. and Jacobs, C., 2023, June. Cyber Warfare and Cyber Terrorism Threats Targeting Critical Infrastructure: A HCPS-based Threat Modelling Intelligence Framework. In ECCWS 2023 22nd European Conference on Cyber Warfare and Security. Academic Conferences and Publishing Limited.

Nkongolo, M., 2023, January. Game Theory based Artificial Player for Morabaraba Game. In 2023 5th International Conference on Smart Systems and Inventive Technology (ICSSIT) (pp. 1210-1218). IEEE.

Nkongolo, M., Tokmak, M. (2023). Zero-Day Threats Detection for Critical Infrastructures. In: Gerber, A., Coetzee, M. (eds) South African Institute of Computer Scientists and Information Technologists. SAICSIT 2023. Communications in Computer and Information Science, vol 1878. Springer, Cham. https://doi.org/10.1007/978-3-031-39652-6_3

Nkopodi, N. and Mosimege, M., 2009. Incorporating the indigenous game of morabaraba in the learning of mathematics. South African Journal of Education, 29(3).

Ortiz-Garces, I., Gutierrez, R., Guerra, D., Sanchez-Viteri, S. and Villegas-Ch, W., 2023. Development of a Platform for Learning Cybersecurity Using Capturing the Flag Competitions. Electronics, 12(7), p.1753.

Stoker, G., Clark, U., Vanajakumari, M. and Wetherill, W., 2021. Building a Cybersecurity Apprenticeship Program: Early-Stage Success and Some Lessons Learned. Information Systems Education Journal, 19(2), pp.35-44.

Traulsen, A. and Glynatsi, N.E., 2023. The future of theoretical evolutionary game theory. Philosophical Transactions of the Royal Society B, 378(1876), p.20210508.

Nkongolo, M., van Deventer, J.P. and Kasongo, S.M., 2022. Using deep packet inspection data to examine subscribers on the network. *Procedia Computer Science*, *215*, pp.182-191.

Nkongolo Wa Nkongolo, M., 2024. RFSA: A Ransomware Feature Selection Algorithm for Multivariate Analysis of Malware Behavior in Cryptocurrency. *International Journal of Computing and Digital Systems*, *15*(1), pp.893-927.

Nkongolo, M. and Sewnath, J., 2024. Data protection psychology using game theory. *arXiv preprint arXiv:2402.07905*.